\documentclass[twoside]{article}

\usepackage{lipsum} 
\usepackage[round]{natbib}
\usepackage[sc]{mathpazo} 
\usepackage[T1]{fontenc} 
\usepackage{microtype} 

\usepackage[hmarginratio=1:1,top=32mm,columnsep=20pt]{geometry} 
\usepackage{multicol} 
\usepackage[hang, small,labelfont=bf,up,textfont=it,up]{caption} 
\usepackage{booktabs} 
\usepackage{float} 
\usepackage{lettrine} 
\usepackage{paralist} 

\usepackage{abstract} 

\usepackage{titlesec} 
\renewcommand\thesection{\Roman{section}} 
\renewcommand\thesubsection{\Roman{subsection}} 
\titleformat{\section}[block]{\large\scshape\centering}{\thesection.}{1em}{} 
\titleformat{\subsection}[block]{\large}{\thesubsection.}{1em}{} 
\usepackage{soul}
\usepackage{footnote}
\usepackage{fancyhdr} 
\usepackage{graphicx}
\DeclareGraphicsExtensions{.jpg}
\usepackage{caption}
\usepackage{subcaption}
\usepackage{amsmath}
\usepackage{booktabs,xcolor,colortbl}
\usepackage{hhline}
\usepackage[hidelinks]{hyperref}

\title{\vspace{-15mm}\fontsize{25pt}{10pt}\selectfont\textbf{Macaque's Cortical Functional Connectivity Dynamics at the Onset of Propofol-Induced Anesthesia}} 

\author{
\large
\textsc{Eduardo C. Padovani}
\thanks{Email: \texttt{eduardo.padovani@alumni.usp.br}}\\[2mm] 
\normalsize \\ 
\normalsize 
\vspace{-5mm}
}
\date{}

\begin{document}

\maketitle 

\thispagestyle{fancy} 


\begin{abstract}

\noindent \ Propofol, when administered for general anesthesia, induces oscillatory dynamic brain states that are thought to underlie the drug’s pharmacological effects. Despite the elucidation of propofol’s mechanisms of action at the molecular level, its effects on neural circuits and overall cortical functioning, which eventually lead to unconsciousness, are still unclear. To identify possible mechanisms, the spatial-temporal patterns of functional connectivity established among specialized cortical areas in anesthetized subjects need to be described. Within this context, the present research involved the analysis of dense sub-dural ECoG electrode array recordings from macaques under propofol anesthetic induction. Granger causality methodology was used to infer functional connectivity interactions in five physiological frequency bands serially over time, every five seconds throughout the experiment. The time-resolved networks obtained permitted us to observe the unfolding of the anesthetic induction and compare the networks obtained under different experimental conditions. About one minute after administering propofol, functional connectivity started to increase for 4-5 minutes, then decreased until the LOC was achieved. A predominant Granger causality flow from the occipital and temporal areas towards the frontal and parietal regions was also evidenced during the transition. During general anesthesia, the local connectivity of the occipital lobe increased, as did the interactions between the occipital and temporal lobes. Conversely, the functional connectivity from the frontal and parietal lobes toward the temporal and occipital regions was mainly impaired. The research presents one of the most comprehensive descriptions of the effects of propofol on large-scale cortical functional connectivity in non-human primates being one of the first studies to describe the dynamics of the functional connectivity during the transitional state that precedes the LOC.
\end{abstract}


\begin{multicols}{2} 

\section{Introduction}
\lettrine[nindent=0em,lines=3]{A}nesthetic agents are neurotropic drugs that revolutionized clinical medicine. Employed for sedation or general anesthesia purposes, they have enabled a series of medical procedures that would otherwise be unimaginable, being among the neurotropic drugs most used all over the world \citep{purdon2015clinical}. Among the available anesthetic agents, propofol (2,6-di-isopropyl phenol) is one of the most widely studied and administered in human patients \citep{purdon2015clinical}, and its pharmacological effects are widely known and described. Furthermore, its molecular mechanisms of action on the modulation of post-synaptic \mbox{GABA-A} receptors, enhancing the inhibition of post-synaptic interneurons and cortical pyramidal neurons, are reasonably understood \citep{bai1999general, hemmings2005emerging, purdon2015clinical}. 
However, as GABA-A receptors are widely present throughout the brain and central nervous system, and given the complexity of the brain, with myriads of interconnections and feedback loops, it is almost impossible to predict the effects on brain dynamics induced by propofol’s modulation of \mbox{GABA-A} receptors \citep{ching2014modeling}. In fact, the effects of propofol on functional connectivity patterns and large-scale functional brain networks are still poorly understood \citep{ishizawa2016dynamics}.
Moreover, propofol anesthetic induction does not seem to follow a unique, uniform unfolding over time. Given that EEG studies in human patients reveal predictable signature changes over the anesthetic induction process \citep{purdon2013electroencephalogram,purdon2015clinical}, it could be hypothesized that distinct neural changes might be happening over time and along with distinct specific cortical areas according to propofol pharmacokinetics and pharmacodynamics \citep{ishizawa2016dynamics}.

It is believed that in order to perform its activities, the brain relies on neural processes involving the joint participation of distinct specialized cortical areas that functionally interact in precise and specific ways \citep{mcintosh2000towards,stam2007graph,tononi1998consciousness}. It is also concerned that these interactions are hierarchically organized; one cortical region might exert influence over another, existing a directed flow of information among them \citep{seth2007distinguishing}. Functional connectivity is also considered to be dynamic. It might reconfigure and present considerable changes at a timescale ranging from 500 ms to a few seconds \citep{tononi1998consciousness,edelman2013consciousness}. 
Nonetheless, it is complicated to study and evaluate these aspects of the working brain experimentally once most of the available recording techniques do not simultaneously provide extensive coverage and high spatial and temporal resolution required to discretize specific cortical regions and their functional connectivity dynamics at their typical time scale \citep{tononi1998consciousness,nagasaka2011multidimensional}.

In the present research, we have analyzed a multi-dimensional recording electrocorticogram (MDR-ECoG) database from the laboratory of adaptive intelligence at the \texttt{RIKEN BRAIN SCIENCE INSTITUTE}, Saitama, Japan. The MDR-ECoG is regarded as the most advanced technology to record cortical neural activity \citep{nagasaka2011multidimensional,yanagawa2013large}, and provides records of (1KHz) of temporal and (5mm) of spatial resolution. The database involved propofol anesthetic induction experiments in two non-human primates subjects of the species \textit{Macaca fuscata} who were chronically implanted with a dense matrix array of \textit{MDR-ECoG} electrodes that continuously covered their left brain hemispheres.

 Aimed at characterizing the alterations in the exchange process of information flow among brain regions throughout distinct phases of propofol anesthetic induction, we have used a methodology based on \textit{Granger causality in the frequency domain} \citep{granger1969investigating,seth2007distinguishing} to infer directed functional interactions among the electrode's time series on five physiological frequency bands. Granger causality interactions were inferred serially over time, at every 5 seconds of the experiment. This methodology allowed us to follow the anesthetic induction unfolding at the typical time scale that changes might happen in the brain. \\

The results reveal a series of alterations in the functional connectivity's spatiotemporal patterns throughout the anesthetic induction process. The administration of a single bolus of propofol (5.0-5.2mg/kg) was followed by a considerable increase in functional connectivity across the five frequency bands analyzed. That period was also characterized by the predominance of the Granger causality flow from the occipital and temporal lobes (causality source) towards the parietal and frontal areas (causality sink). After about 3-4 minutes, the mean functional connectivity began to decrease for 5-6 minutes until the loss of consciousness point (LOC) was reached; subsequently, the mean global functional connectivity remained approximately the same while the macaques were under general anesthesia. During general anesthesia, there was an increase in the occipital lobe's local functional connectivity in the five frequency bands analyzed. Functional connectivity between the occipital and temporal lobes also increased in high-frequency bands (Beta and Gamma). Granger causality interactions, leaving the parietal lobe toward the frontal lobe, raised in the Theta and Alpha frequency bands. Conversely, functional connectivity from the parietal lobe towards the occipital lobe decreased in the Delta, Theta, and Alpha frequency bands. A general impairment was verified in the Granger causality interactions leaving the frontal and parietal lobes towards the occipital and temporal regions.

\section{Methods}

The present research was based on the analysis of an ECoG electrophysiological neural record database. The database came from the Laboratory of Adaptive Intelligence at the \texttt{RIKEN BRAIN SCIENCE INSTITUTE}, Saitama, Japan. Experimental, surgical, and recording procedures were performed by researchers from the respective institution according to the experimental protocols (No. H24-2-203(4)) approved by the \texttt{RIKEN} ethics committee and the recommendations of the Weatherall report, "The use of non-human primates in research." For further information regarding methodology, subjects, and materials, see \citep{nagasaka2011multidimensional} and (\texttt{http://neurotycho.org}).
The analyzed databases are relative to two adult male macaques of the species \textit{Macaca fuscata}. Both animal models had an MDR-ECoG electrode array chronically implanted in the subdural space of the left brain hemisphere. One hundred and twenty-eight ECoG electrodes with an average inter-electrode distance of 5 mm covered the frontal, parietal, and occipital lobes. In addition, electrodes were also implanted in the frontal, parietal, and occipital medial walls.
The database comprises four propofol anesthetic induction experiments conducted over different days. Two experiments were performed on monkey A (male) and two on monkey B (male).
In each experiment, the neural activity was recorded for approximately ten minutes with macaques in awake conditions with their eyes covered. Then, a single bolus of propofol (5.2 mg/Kg monkey A and 5.0 mg/Kg monkey B) was injected intravenously to induce anesthesia. The loss of consciousness point was clinically set, based on their lack of response to external stimuli, such as touching the nostrils or opening the hands. After achieving LOC, neural activity was recorded for approximately ten minutes.

\subsection*{Data Analysis and Experimental Procedures :}

The estimation of functional neural networks was performed according to the following steps:

\begin{itemize}

\item Databases of MDR-ECoG electrophysiological neural activity recordings were used.

\item Each electrode of the matrix array was considered a node of the network and resembled the cortical region in which it was positioned.

\item Granger causality in the frequency domain was used as a neural connectivity estimator to infer functional connectivity interactions among the electrode's records (time series).

\item Pairwise association values between the electrodes were saved into adjacency matrices.

\item Matrices were estimated serially over time at every five seconds of the neural recording experiment in five physiological frequency bands.

\end{itemize}

\subsection{Signal Processing and Granger Causality in the Frequency Domain:}

\begin{enumerate}

\item An IIR-notch filter was applied to attenuate components of the signal at 50Hz.

\item The signal was down-sampled from 1KHz to 200Hz.

\item The electrophysiological time series were divided into windows of 1000 points, equivalent to a five-second recording of neural activity.

\item On each time series, the trend was removed, and the average was subtracted.

\item The tests KPSS \citep{kwiatkowski1992testing} and ADF \texttt{[Augmented Dickey Fuller]} \citep{hamilton1989new} were applied to verify the time series stationary conditions.

\end{enumerate}

\subsection{Computation of Causal Interactions}

For the computation of Granger causality in the frequency domain functional interactions, with some adaptations, the following libraries were used: \texttt{MVGC GRANGER TOOLBOX} \citep{seth2010matlab} and the library \texttt{BSMART toolbox} [\textit{\textbf{B}rain-\textbf{S}ystem for \textbf{M}ultivariate \textbf{A}uto\textbf{R}egressive \textbf{T}imeseries} \texttt{toolbox}] \citep{cui2008bsmart}.

\begin{enumerate}

\item \textbf{Model Order}
  
 The criteria of model selection from Akaike (AIC) and Bayes/Schwartz (BIC) were used to infer the model order. Both methods returned a model order parameter equal to seven.

\item \textbf{Causal Interactions}

At each window of 1000 points, Granger causality in the frequency domain interactions was pair-wise computed among the 128-time series by the use of the function \texttt{cca\_pwcausal()} (\texttt{MVGC GRANGER TOOLBOX}) in five physiological frequency bands: Delta (0-4Hz), Theta (4-8Hz), Alpha (8-12Hz), Beta (13-30Hz) and Gamma (25-100Hz). The interaction values obtained were saved in adjacency matrices.

\end{enumerate}

\subsection*{T-Distributed Stochastic Neighbor Embedding Plots}

The functional connectivity matrices serially estimated over time along with the anesthetic induction experiment were projected into a bi-dimensional representation using a \textit{t-Stochastic Neighborhood Embedding algorithm} \citep{van2008visualizing}. The respective entries vectors of the weighted directed functional connectivity matrices were used as input features into the t-SNE algorithm \citep{van2008visualizing} implemented in the R-CRAN package Rtsne \citep{krijthe2018package}, with the following parameters: perplexity=30, exaggeration factor=12, and the maximum number of interactions=500.

\subsection*{Brain Functional Networks}

To obtain directed unweighted networks, for each sequence of graphs respective to a physiological frequency band, a threshold was applied, and only interactions with magnitude values higher than the threshold were considered edges of the graphs. The threshold was set to 5\% higher interaction values of all matrices of the sequence corresponding to each frequency band. This procedure finally resulted in directed, unweighted functional brain networks.

To ensure that our results were not exclusively dependent on the chosen threshold, twelve distinct values of threshold respective from 0.5\% to 15\% higher interaction values were evaluated (0.5\%, 1\%, 2\%, 3\%, 4\%, 5\%, 6\%, 7\%, 8\%, 9\%, 10\%, and 15\% ). The same dynamic behavior on the network's average degree over time across the anesthetic induction experiment was verified along with the different thresholds; see complementary material\footnote{Complementary material is not currently available online.}.
The manuscript shows results respective to one of the four experiments performed; the other three experiments were used to validate the results presented. See complementary material to compare the results and figures of all four independent experiments.

\subsection{Functional Connectivity Alterations Due to Distinct Experimental Conditions}

The present research investigated unweighted-directed large-scale functional brain networks obtained by applying a threshold over the Granger causality association matrices. The analysis of functional connectivity and Granger causality flow among brain areas was evaluated in terms of the number of connections in these directed binary networks. Thus, the analysis did not differentially take into account the exact magnitude of association values returned by the Granger causality method. 
The anesthetic induction experiments in the macaque animal models involved three distinct conditions: alert with eyes closed, transition, and general anesthesia. To describe and verify the alterations due to the administration of propofol in the brain state, we compared the populations of functional brain networks respective to the transition and general anesthesia phases to those estimated during the resting state with eyes closed conditions, a control state without the presence of propofol, which supposedly reflects natural brain activity in the absence of cognitive demands and visual-evoked responses.

Once the recording technique and methodology applied provided several networks throughout the experiment, populations of networks respective to each stage were considered. Therefore, it permitted us to make quantitatively legitimate comparisons among the different physiological states. Properties were evaluated under a statistical test to verify whether significant distinctions occurred or not due to the administration of propofol throughout the transition and general anesthesia conditions. 
For each property evaluated, the Wilcoxon signed-rank test with a 5 percent p-value was independently applied in each of the four experiments on both conditions [greater] and [smaller]. In the manuscript, only alterations confirmed by the statistical test simultaneously in both four independent realized experiments were considered and reported as results.

\section{Results}

The present research observed a range of results and phenomena after administering a single bolus (5.0-5.2 mg/Kg) of the anesthetic agent propofol to the macaque animal models. Being verified alterations in the patterns and dynamics of functional brain connectivity.


\end{multicols}

\begin{figure}[!h]
\begin{subfigure}{.5\textwidth}
  \centering
  \includegraphics[width=1\linewidth]{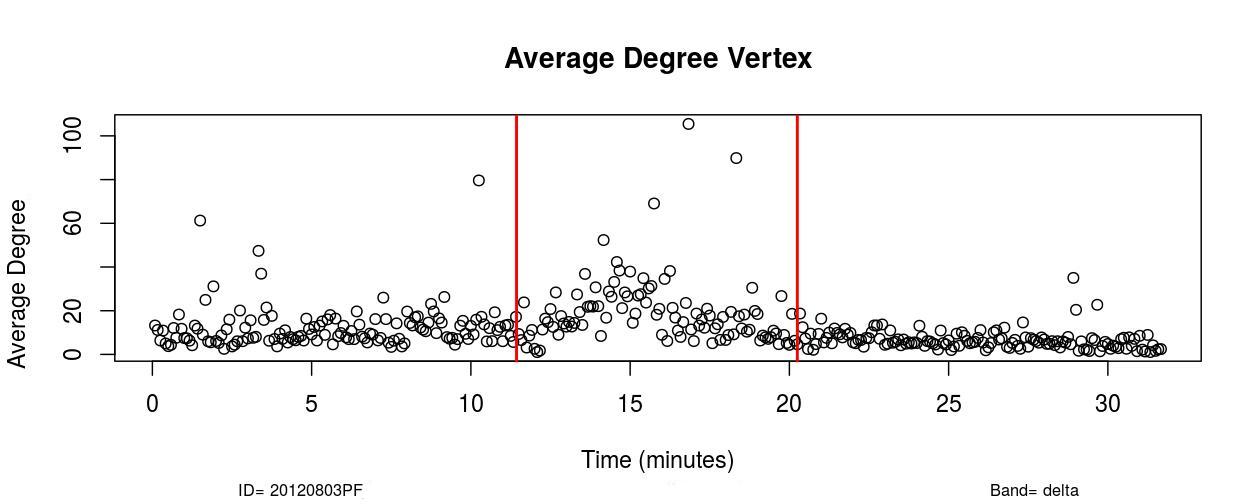}
  \caption{Delta 0-4Hz}
  \label{fig:sfig1}
\end{subfigure}%
\begin{subfigure}{.5\textwidth}
  \centering
  \includegraphics[width=1\linewidth]{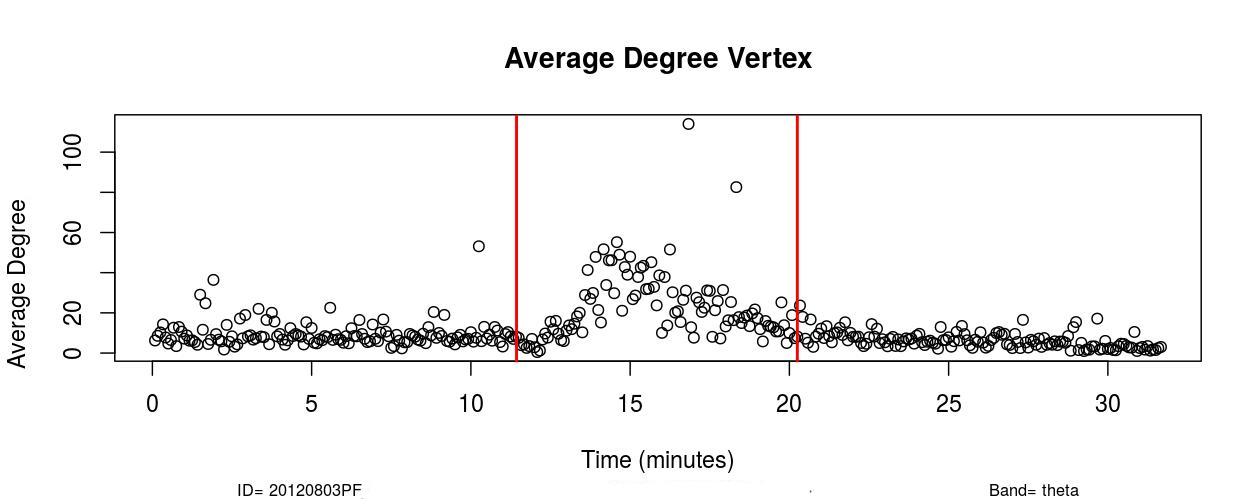}
 \caption{Theta 4-8Hz}
  \label{fig:sfig2}
\end{subfigure}\\
\centering
\begin{subfigure}{.5\textwidth}
\includegraphics[width=1\linewidth]{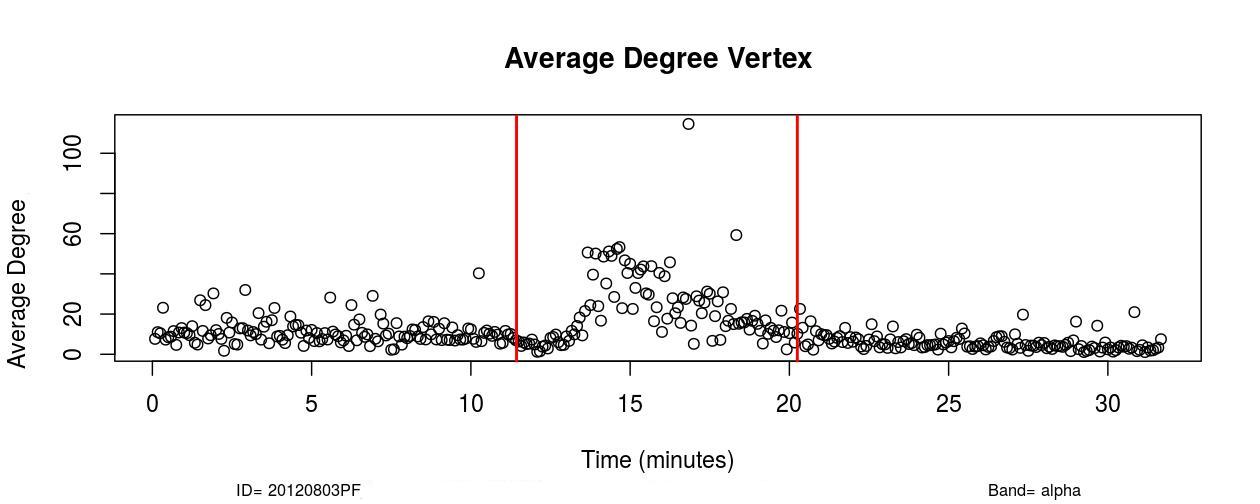}
  \caption{Alpha 8-12Hz}
  \label{fig:sfig3}
\end{subfigure}%
\begin{subfigure}{.5\textwidth}
  \centering
  \includegraphics[width=1\linewidth]{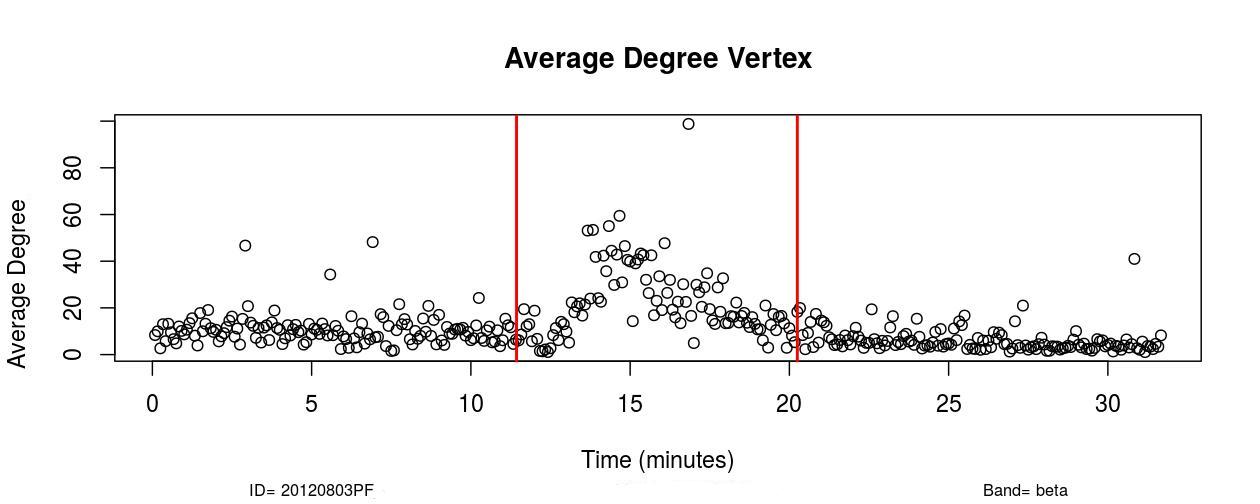}
  \caption{Beta 13-30Hz}
  \label{fig:sfig4}
\end{subfigure}\\
\centering
\begin{subfigure}{.5\textwidth}
\includegraphics[width=1\linewidth]{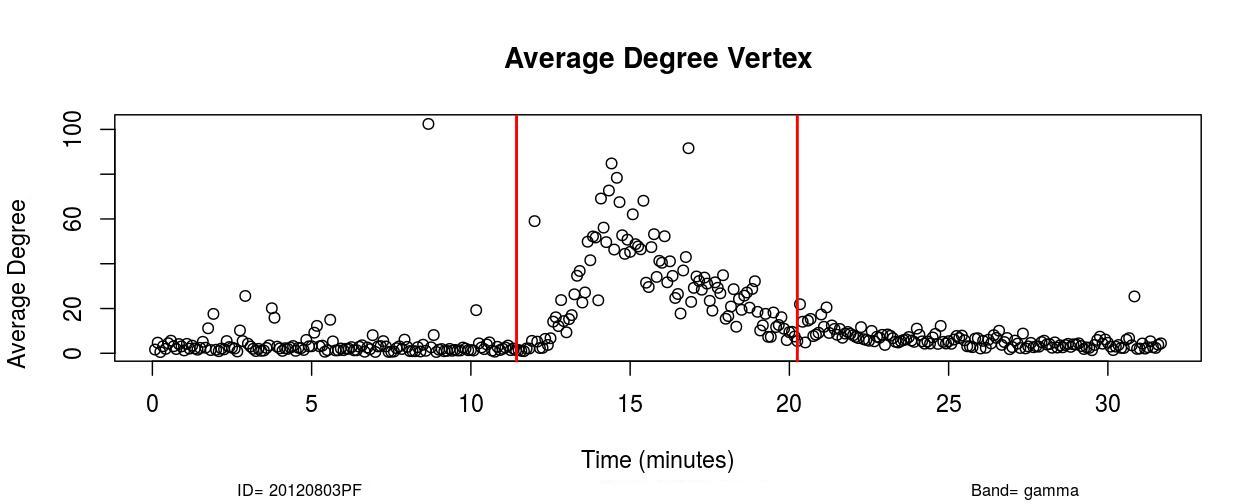}
  \caption{Gamma 25-100Hz}
  \label{fig:sfig3}
\end{subfigure}%
\caption{\textbf{Average functional connectivity increased after the administration of propofol}. Average node degree over time during propofol anesthetic induction on the physiological frequency bands A) Delta (0-4Hz), B) Theta (4-8Hz), C) Alpha (8-12Hz), D) Beta (13-30Hz), and E) Gamma (25-100Hz). In each Sub-Figure, the first vertical red line (taking place after 11-12 minutes the experiment began) indicates the time when the anesthetic was administered; the second red line (21 minutes) indicates the time when the loss of consciousness point (LOC) took place. Within the period from 0 to 11 minutes, the macaque was awakened with eyes closed; from 21 to 34 minutes, the macaque was in general anesthesia. The administration of the anesthetic agent propofol induced an increase in functional connectivity for about 3-4 minutes in the five frequency bands analyzed. Then connectivity gradually decreased until LOC was achieved and remained nearly constant while the animal was anesthetized.}
\label{fig:fig}
\hypertarget{FIGURE1}{}
\end{figure}

\begin{multicols}{2}

\subsection{Profofol First Induces a Decrease in Functional Connectivity:}

The first remarkable alterations in functional connectivity occurred within 20-25 seconds after propofol administration when the average degree of functional brain networks diminished for about 20-25 seconds. As it is possible to observe in \hyperlink{FIGURE1}{$Figure \cdot 1$}, a decrease in the mean functional connectivity occurred on the Delta (0-4Hz), Theta (4-8Hz), Alpha (8-12Hz), and Beta (13-30Hz) frequency bands almost right after the administration of propofol, which is indicated by the first red vertical line, time=11 minutes (see \hyperlink{FIGURE1}{$Figure \cdot 1$, \mbox{\textit{Sub-Figures 1 to 4}}}). The same conclusions can be drawn by analyzing \hyperlink{FIGURE2}{$Figure \cdot 2$}, where it is possible to verify the respective period when functional connectivity decreased (see \hyperlink{FIGURE2}{$Figure \cdot 2$, \mbox{\textit{Sub-Figures 12 to 16}}}). As each Sub-Figure corresponds to 5 seconds of neurophysiological records, it is evidenced that this first decrease in functional connectivity lasted for about 20-25 seconds. The results shown in \hyperlink{FIGURE3}{$Figure \cdot 3$} reveal large areas of the cortex, mainly over the occipital lobe, colored in white, thus demonstrating that this cortical region did not have a resultant Granger causality flow, supposedly due to the absence or reduced number of connections. This result also corroborates the conclusions obtained from Figures 1 and 2.

\subsection{Increase in Functional Connectivity Preceding LOC}

 Following the decrease in functional connectivity that occurred within 25 seconds after the drug administration, functional connectivity then started to gradually increase for about 4-5 minutes, achieving a maximum peak and later started to progressively decrease for about 6-7 minutes until the loss of consciousness point (LOC) was achieved. From \hyperlink{FIGURE1}{$Figure \cdot 1$}, it is verified that this increase in functional connectivity occurred in the five physiological frequency bands analyzed, and the most pronounced increase occurred in the frequency band Gamma (25-100Hz). By analyzing \hyperlink{FIGURE2}{$Figure \cdot 2$}, one can verify on the frequency band Beta (13-30 Hz) how this phenomenon of gradual increase in functional connectivity manifested on individual nodes over brain areas (see \hyperlink{FIGURE2}{$Figure \cdot 2$}, \mbox{\textit{Sub-Figures 20-35}}), first 80 seconds of the gradual increase. From (\hyperlink{FIGURE2}{$Figure \cdot 2$, \mbox{\textit{Sub-Figures 20-35}}}), it was also verified that almost all electrodes were highly connected, revealing that this high connectivity effect involved all cortical regions in which electrodes were positioned.

\end{multicols}

\begin{figure*}[!h]
  \includegraphics[width=\textwidth,height=12cm]
  {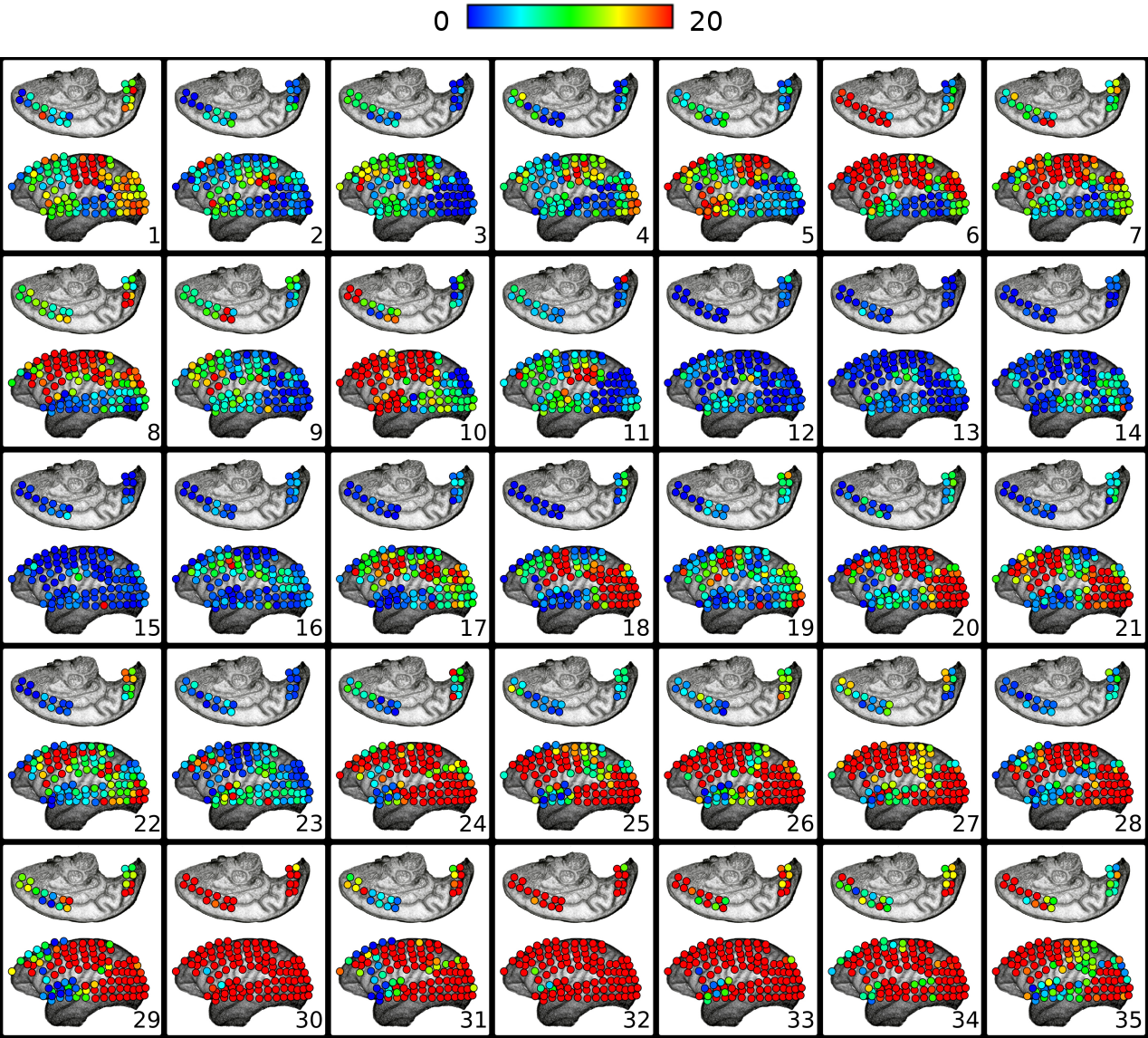}
     \caption{\textbf{Functional connectivity over space and time after propofol administration - Beta (13-30Hz)}. In each Sub-Figure,  the electrode's positions are represented over the respective brain areas, and a color gradient indicates the degree of each node. The Sub-Figures were estimated sequentially over time and are shown respectively in subsequent 5 seconds intervals. Propofol was administered at the time corresponding to Sub-Figure 6. After twenty-five seconds of the drug administration, a reduction in functional connectivity occurred, lasting for about 20-25 seconds, see Sub-Figures (12-16). Subsequently, functional connectivity gradually increased, see Sub-Figures (17-25) until achieving a state in which the majority of nodes were highly connected, see Sub-Figures (25-35).
}
\hypertarget{FIGURE2}{}
\end{figure*}

\begin{multicols}{2}


\subsection{Granger Causality Resultant Flow Profile over Brain Areas Along with the Anesthetic Induction}

Granger Causality is a functional connectivity estimator aimed at inferring the flow of information from one cortical area to another. As a measure that offers the direction in which the functional interaction occurs, the resultant OUT-DEGREE minus the IN-DEGREE of each node can be interpreted as a measure of causal flow action associated with the respective node. This resultant causal flow action profile identifies at a given moment the nature of preferentially affecting (assuming a role of a source of Granger causality) or being affected by other nodes (assuming an action as a sink of Granger causality). Therefore, provide an estimate of the respective cortical area in preferentially sending or receiving information from other cortical regions. 
\hyperlink{FIGURE3}{$Figure \cdot 3$} displays the resultant causal flow profile of the electrodes over their respective cortical positional coordinates along time, at every five seconds, right after propofol administering. From \hyperlink{FIGURE3}{$Figure \cdot 3$}, one can notice that when the brain entered the regime characterized by high connectivity, which was established about 110 seconds after the drug injection (see \hyperlink{FIGURE3}{$Figure \cdot 3$, \mbox{\textit{Sub-Figures 22-35}}}),
electrodes correspondent to the occipital and temporal lobes had a high positive causal flow. This result reveals that these brain areas tended to exert a strong causal influence and assumed predominant action as a source of Granger causality. On the other hand, electrodes of the parietal and frontal lobes predominantly had a negative resultant causal flow profile. Nodes with negative causal flow reflect on cortical areas predominantly affected by other regions. The obtained results, therefore, evidenced that areas of the frontal and parietal lobes acted mainly as sinks of Granger causality during the transition. 

This pattern took place during the transition state when functional connectivity significantly increased over the entire cortex. Once presented as one main recurrent pattern, as opposed to resting-state conditions in which the brain presented a much more dynamic and diverse repertoire of states, it showed that the brain entered into a very particular state in which cortical areas assumed one predominant definite behavior of the Granger causality flow dynamics.

\subsection{Large-Scale Exchange Granger Causality Flow Among Cortical Lobes}

To verify the alterations caused by the administration of propofol on the exchange process of Granger causality flow among the cortical brain lobes, for each time-resolved functional network, we have counted the number of directed connections leaving one cortical lobe to another on all combinations (frontal → parietal, frontal → temporal, frontal → occipital, parietal → frontal, parietal → temporal, parietal → occipital, temporal → frontal, temporal → parietal, temporal → occipital, occipital → frontal, occipital → parietal and occipital→temporal). Once the average degree varied considerably throughout the experiment, we then normalized it by the average degree of each time-resolved network. Then we compared the network populations of the transition and general anesthesia phases to the resting state networks. Finally, by applying the Wilcoxon signed-rank test with a 5 percent p-value on both conditions [greater] and [smaller], we reported the results, which were confirmed by the tests concurrently in all four experiments independently realized.

\begin{itemize}

\item Transition Phase

When compared to the resting state conditions, the main alterations observed were increased interactions from inferior-posterior (temporal and occipital lobes) to anterior parts of the brain (frontal and parietal lobes), see (\hyperlink{FIGURE4}{$Figure \cdot 4$}, \mbox{\textit{Sub-Figures A, B, C, and D}}).
 
In the Delta (0-4Hz) frequency band, it was verified that there was an increase in the interactions from the occipital lobe to the frontal and parietal lobes and also from the temporal lobe toward the parietal lobe (see \hyperlink{FIGURE4}{$Figure \cdot 4$, \mbox{\textit{Sub-Figure A}}}). In Theta (4-8Hz), interactions from the occipital lobe to the frontal and parietal lobes and also interactions from the temporal lobe to the frontal lobe increased (see \hyperlink{FIGURE4}{$Figure \cdot 4$, \textit{\mbox{Sub-Figure B}}}). In Alpha (8-12Hz), an increase in the number of interactions from the temporal lobe towards the frontal and parietal lobes and from the occipital lobe to the frontal lobe occurred (see \hyperlink{FIGURE4}{$Figure \cdot 4$}, \mbox{\textit{Sub-Figure C}}). Finally, in Beta (13-30Hz), an increase in the interactions was verified from the occipital lobe toward the frontal and parietal lobes and from the temporal lobe to the parietal lobe (see \hyperlink{FIGURE4}{$Figure \cdot 4$, \mbox{\textit{Sub-Figure D}}}).
 It was not verified any interactions among brain lobes that statistically significantly decreased during the transition phase.

\newpage

\item General Anesthesia Phase

In general, interactions between the occipital and temporal lobes increased in both directions in the higher frequency bands Beta (13-30Hz) and Gamma (25-100Hz). Connectivity from the parietal to the frontal lobe also increased in the Theta (4-8Hz) and Alpha (8-12Hz) frequency bands. Conversely, the interactions from the parietal lobe to the occipital lobe decreased in the lower and medium frequency bands.
In the Delta (0-4Hz) frequency band, there was a decrease in the interactions from the parietal lobe to the occipital lobe (see \hyperlink{FIGURE4}{$Figure \cdot 4$, \mbox{\textit{Sub-Figure F})}}. In the Theta (\mbox{4-8Hz}), an increase in interactions from the occipital lobe to the frontal lobe and from the parietal lobe to the frontal lobe occurred. However, functional connectivity from the parietal to the occipital lobe diminished (see \hyperlink{FIGURE4}{$Figure \cdot 4$, \mbox{\textit{Sub-Figure G}})}. In the Alpha (8-12Hz) frequency band interactions increased from the parietal areas to the frontal lobe. However, there was a decrease in connectivity from the frontal to the occipital lobe and from the parietal to the occipital lobe (see \hyperlink{FIGURE4}{$Figure \cdot 4$, \mbox{\textit{Sub-Figure H}}}). In the Beta (\mbox{13-30Hz}), a decrease in the interactions from the frontal to temporal lobe was verified, and an increase in the interactions from the temporal lobe to the occipital lobe and from the occipital regions to the temporal areas was also observed (see \hyperlink{FIGURE4}{$Figure \cdot 4$, \textit{\mbox{Sub-Figure I}})}. Finally, in the Gamma (\mbox{25-100Hz}), a decrease in functional connectivity from the temporal lobe to the parietal lobe. Additionally, an increase in the number of interactions from the temporal lobe to the occipital lobe and from the occipital lobe to the temporal lobe was also verified (see \hyperlink{FIGURE4}{$Figure \cdot 4$, \textit{\mbox{Sub-Figure H}}}).
Particularly different from the transition phase, where only an increase in interactions was verified, during the general anesthesia state, a decrease in certain brain functional connections established among brain lobes was verified on all frequency bands analyzed. The functional connectivity interactions that were most compromised were those from the anterior (fronto-parietal) to the inferior posterior (temporal-occipital) cortical areas.
 
\end{itemize}

\subsection*{Considerations Transition State:}

The transitional state that occurred between the resting-awake and general anesthesia conditions, which are regarded as two physiologically stable and coherent states, did not demonstrate that it took place as a single-step change or as a continuous unraveling of one single neural process. Instead, functional connectivity first decreased considerably, then started to increase for a few minutes, and then later decreased again until LOC was achieved. This succession of changes in functional connectivity dynamics seemed consistent with the unfolding of a series of different neural processes.

\subsection{Evaluation of Functional Connectivity at Distinct Levels:}

Once the MDR-ECoG technique provided at the same time, great coverage of the cortical surface along with high spatial resolution permitted us to evaluate functional connectivity at distinct levels.


\end{multicols}

\begin{figure*}[!ht]
  \includegraphics[width=\textwidth,height=14cm]
  {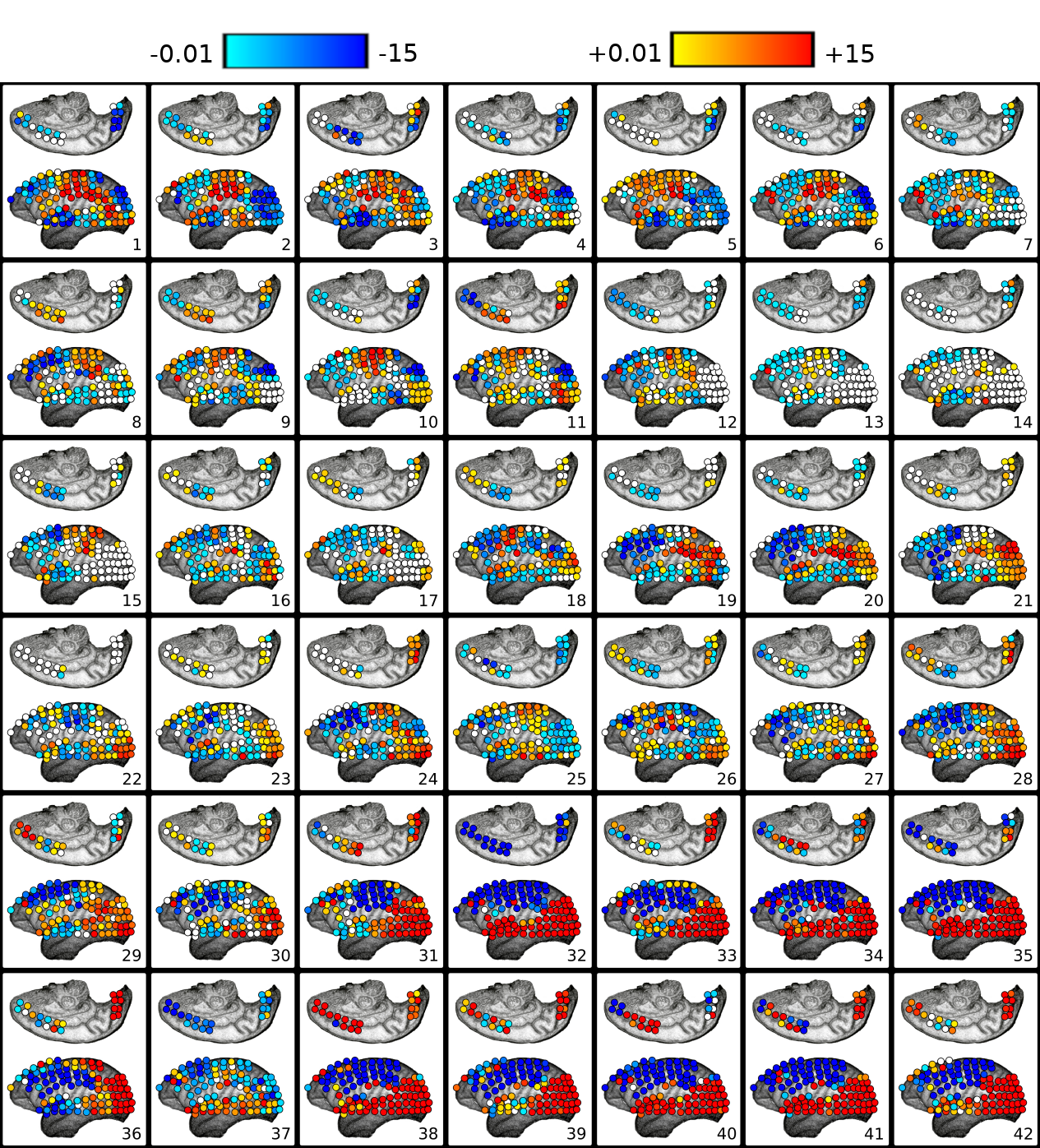}
     \caption{\textbf{Granger causality resultant flow profile over space and time after propofol administration  \mbox{- Alpha (8-12Hz)}}. In each Sub-Figure, the positions of the electrodes are shown in their respective brain areas. The Sub-Figures were estimated sequentially over time and are shown respectively in subsequent 5 seconds intervals. The color gradient indicates the resultant flow (OUT-DEGREE minus the IN-DEGREE) action of each node as a source or sink of Granger causality. Propofol was administered at the time corresponding to Sub-Figure 7. Within about 20-25 seconds after the drug administration, large areas of the occipital lobe and posterior areas of the temporal lobe did not present a resultant causal flow profile, once characterized by white color. After 100-110 seconds from the injection, the cortex causal flow dynamics assumed a predominant pattern over time. The action of each cortical position in the occipital and temporal lobes as a source of causality is indicated in red, and the action of the frontal and parietal areas as a sink of causality is indicated in blue.
}
\hypertarget{FIGURE3}{}
\end{figure*}

\begin{multicols}{2}


Functional connectivity at the global level was evaluated to ascertain how the functional connectivity of the whole large-scale cortical networks changed according to the distinct experimental conditions. We evaluated the average degree of the complete network (all nodes and all connections). Reflected on how connected or disconnected the brain networks were in each phase of the experiment.

\begin{itemize}

\item Transition Phase

 The Wilcoxon test with a p-value of 5\% confirmed that, during the transition, the global functional connectivity increased on all frequency bands analyzed in both four experiments.
\enlargethispage{\baselineskip}\\

In Delta (0-4Hz), the average connectivity in alert conditions was about 12; during the transition, it increased and achieved a peak by order of 50 (see \hyperlink{FIGURE1}{$Figure \cdot 1$, \textit{\mbox{Sub-Figure A}}}.
In Theta (4-8Hz), Alpha (8-12Hz), and Beta (13-30Hz), the average connectivity in the resting state was on the order of 10-12, and during the transition, it achieved a peak of around 60. This observation reveals that functional connectivity increased up to six times during the transition at those frequency bands (see \hyperlink{FIGURE1}{$Figure \cdot 1$, \textit{\mbox{Sub-Figures B, C, and D}})}.

In the Gamma frequency band (\mbox{25-100Hz}), the average degree was about 5-6 and increased to around 80 during the peak, thus verifying that functional connectivity increased up to about 15 times in the Gamma frequency band (see \hyperlink{FIGURE1}{$Figure \cdot 1$, \textit{\mbox{Sub-Figure E}})}.

\item General Anesthesia Phase

The statistical test confirmed that, during general anesthesia, functional connectivity at the global level increased in the Gamma frequency band (25-100Hz). However, on the frequency bands Delta (0-4Hz), Theta (4-8Hz), Alpha (8-12Hz), and Beta (13-30Hz), the statistical tests were inconclusive. This result is very interesting once it demonstrates that propofol-induced general anesthesia is not necessarily given by means of a substantial decrease in cortical functional connectivity at global levels (see \hyperlink{FIGURE1}{$Figure \cdot 1$}).

\end{itemize}

\subsubsection{Brain Lobe's Functional Connectivity at Local Level:}

The brain lobe's functional connectivity at the local level was evaluated to verify how the internal functional connectivity of each brain lobe (frontal, parietal, temporal, and occipital) was affected by the anesthetic agent propofol during the transition and general anesthesia phases. This measure only takes into account the nodes belonging to the respective brain lobe and the connections established among them (intra-lobe connections). Reflects on how cohesive or dispersed the functional connectivity of the particular brain lobe is at a local level.

\begin{itemize}

\item Transition Phase

In the Delta (0-4Hz) frequency band, an increase in functional connectivity was verified over the frontal and temporal lobes.
Functional connectivity increased at the parietal lobe in the Theta (4-8Hz) and Beta (13-30Hz) frequency bands.
In Alpha (8-12Hz), the statistical test was inconclusive for all brain lobes analyzed. 
In Gamma (25-100Hz) Wilcoxon statistical test indicated that functional connectivity at the local level increased in the frontal, parietal, temporal, and occipital lobes. 

\item General Anesthesia Phase

At the local level, the statistical test confirmed that functional connectivity at the local level increased in the occipital lobe on all frequency bands. However, in Theta (4-8Hz) and Alpha (8-12Hz), functional connectivity decreased in the parietal lobe, and in Beta (13-30Hz), functional connectivity decreased in the temporal lobe.

\end{itemize}


\end{multicols}

\begin{figure}[htb]
\centering
  \begin{subfigure}[b]{.19\linewidth}
    \centering
    \includegraphics[width=.99\textwidth]{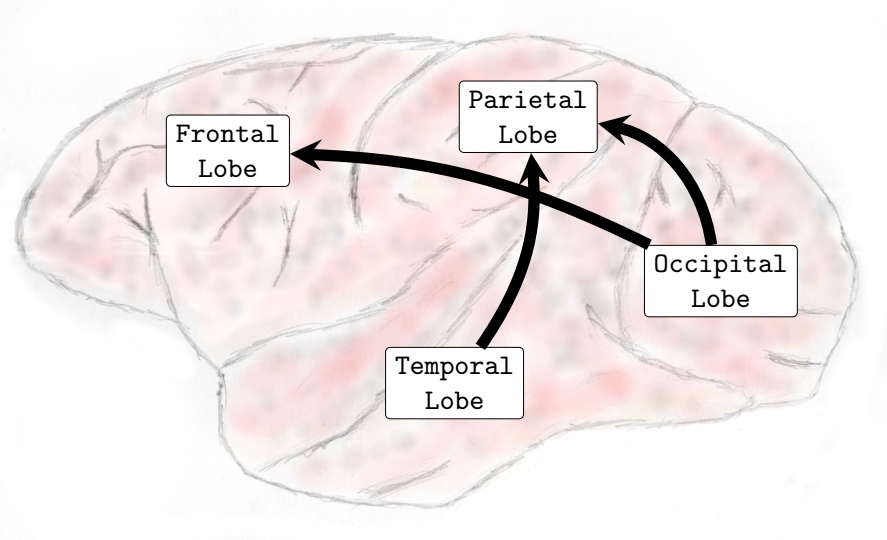}
    \caption{Transition \newline    Delta (0-4Hz)}\label{fig:1a}
  \end{subfigure}%
  \begin{subfigure}[b]{.19\linewidth}
    \centering
    \includegraphics[width=.99\textwidth]{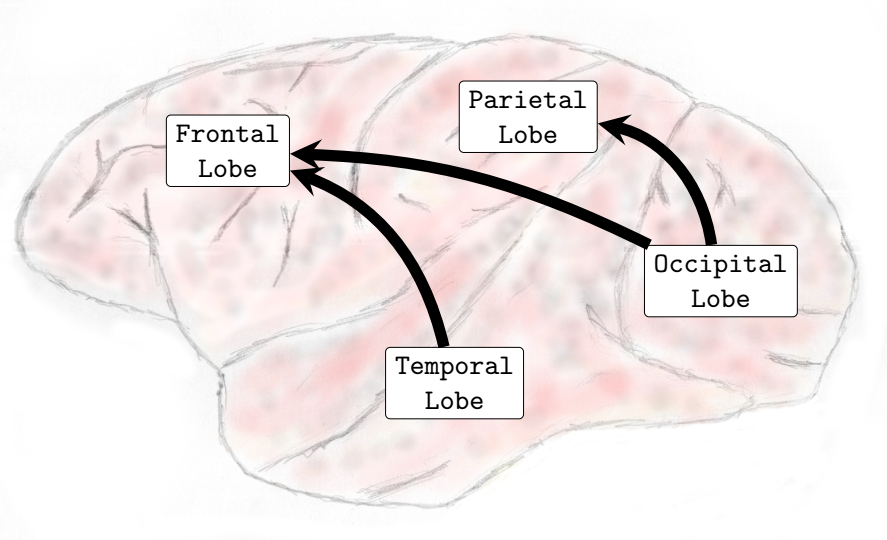}
    \caption{Transition \newline Theta (4-8Hz)}\label{fig:1b}
  \end{subfigure}%
  \begin{subfigure}[b]{.19\linewidth}
    \centering
    \includegraphics[width=.99\textwidth]{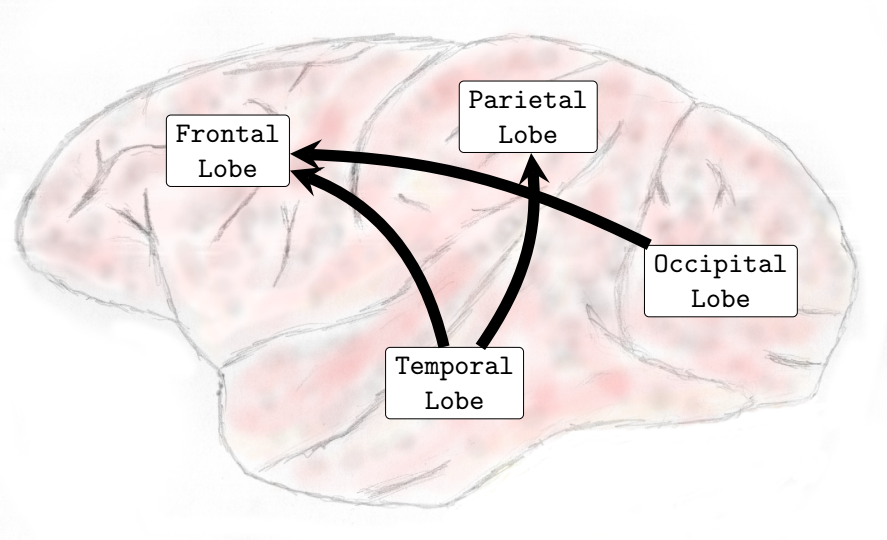}
    \caption{Transition \newline Alpha (8-12Hz)}\label{fig:1c}
  \end{subfigure}%
  \begin{subfigure}[b]{.19\linewidth}
    \centering
    \includegraphics[width=.99\textwidth]{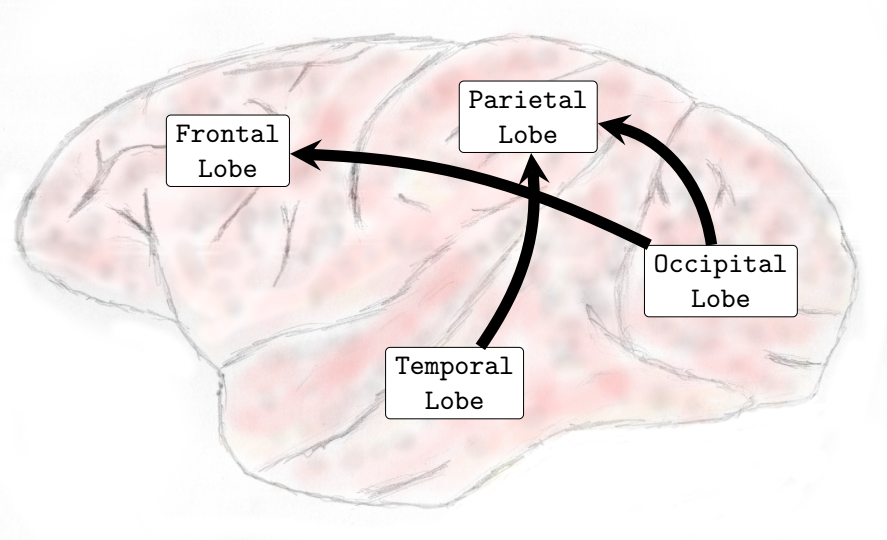}
    \caption{Transition \newline   Beta (13-30Hz)}\label{fig:1d}
  \end{subfigure}%
  \begin{subfigure}[b]{.19\linewidth}
    \centering
    \includegraphics[width=.99\textwidth]{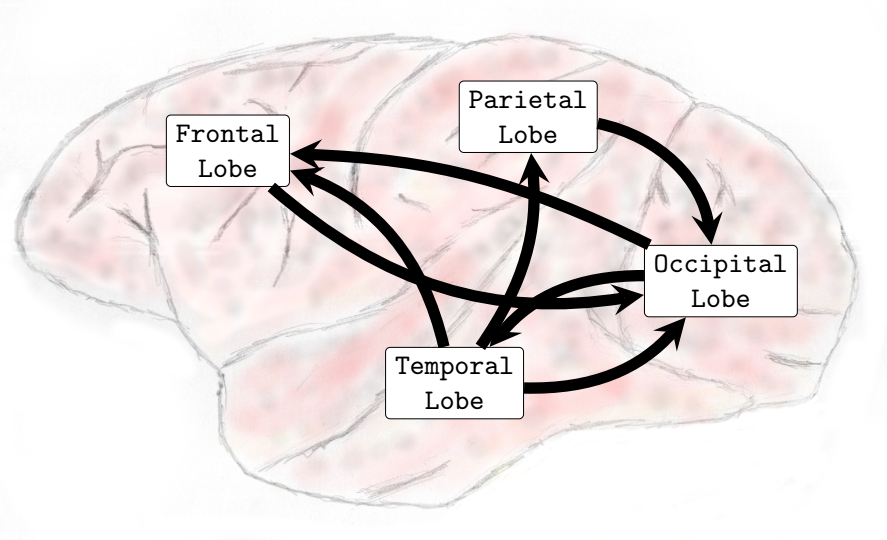}
    \caption{Transition \newline  Gamma (25-100Hz)}\label{fig:1e}
  \end{subfigure}\\%
  \begin{subfigure}[b]{.19\linewidth}
    \centering
    \includegraphics[width=.99\textwidth]{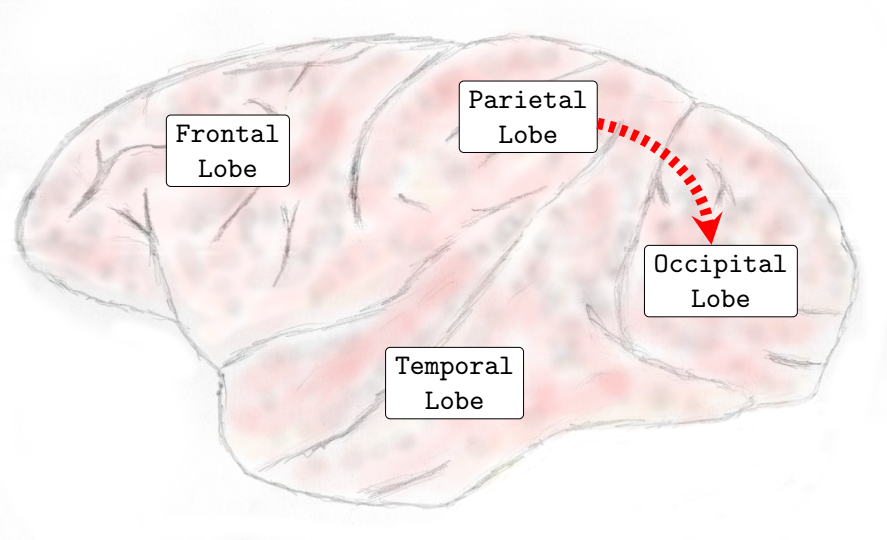}
    \caption{Anesthesia \newline Delta (0-4Hz)}\label{fig:1f}
  \end{subfigure}%
  \begin{subfigure}[b]{.19\linewidth}
    \centering
    \includegraphics[width=.99\textwidth]{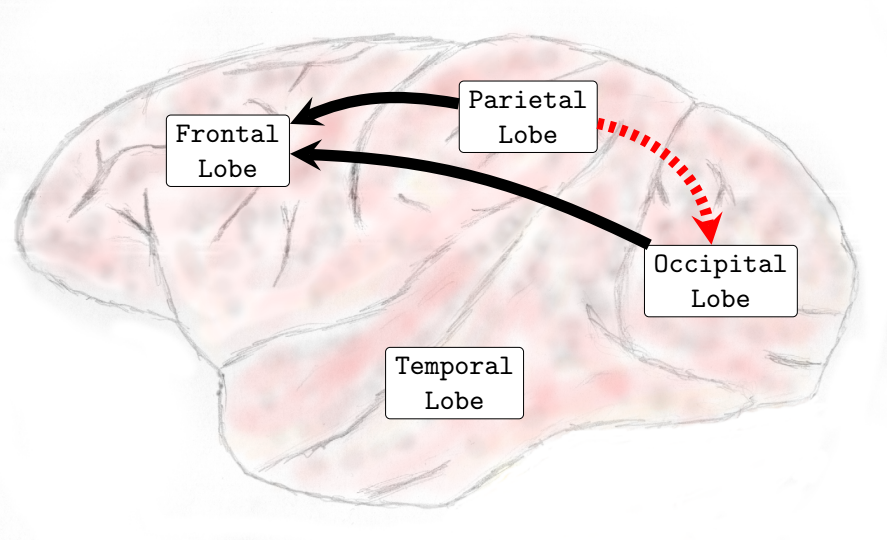}
    \caption{Anesthesia \newline Theta (4-8Hz)}\label{fig:1g}
  \end{subfigure}%
  \begin{subfigure}[b]{.19\linewidth}
    \centering
    \includegraphics[width=.99\textwidth]{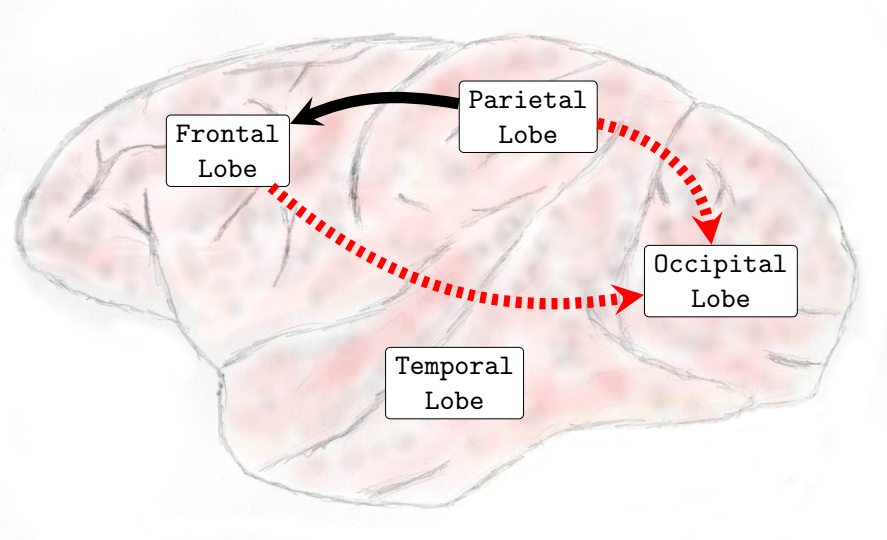}
    \caption{Anesthesia \newline Alpha (8-12Hz)}\label{fig:1h}
  \end{subfigure}%
  \begin{subfigure}[b]{.19\linewidth}
    \centering
    \includegraphics[width=.99\textwidth]{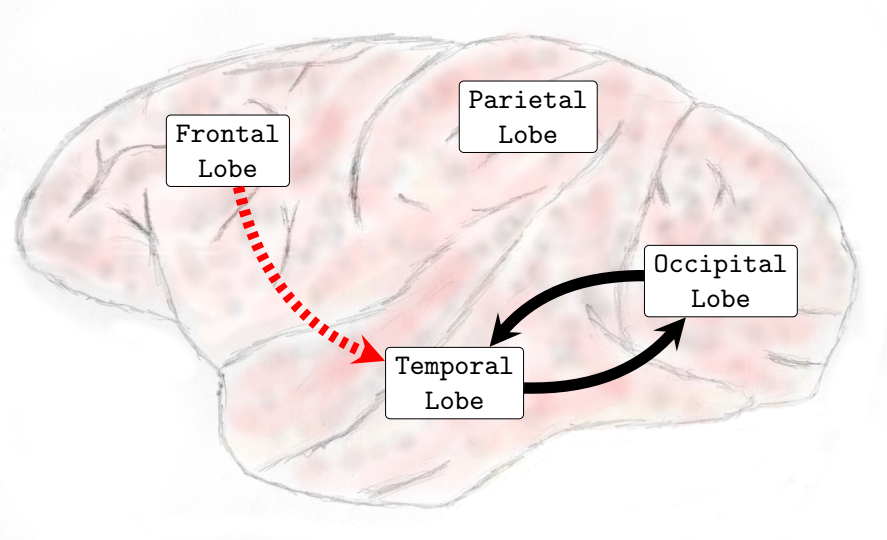}
    \caption{Anesthesia \newline  Beta (13-30Hz)}\label{fig:1i}
  \end{subfigure}%
  \begin{subfigure}[b]{.19\linewidth}
   \centering
    \includegraphics[width=.99\textwidth]{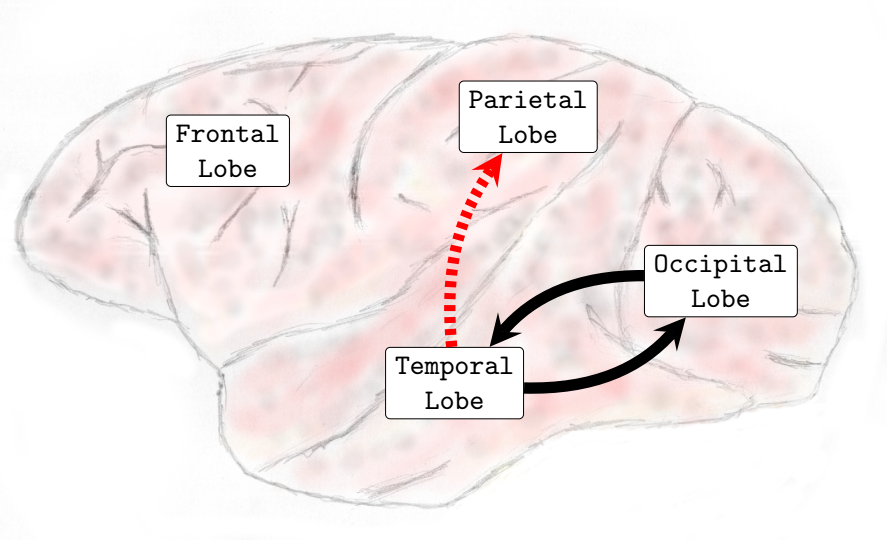}
    \caption{Anesthesia \newline Gamma (25-100Hz)}\label{fig:1j}
  \end{subfigure}%
  \caption{\textbf{Alteration in the connectivity established among brain lobes}. In this figure are represented functional connectivity interactions established among brain lobes which statistically significantly increased or decreased concerning the resting state phase, during the transition phase (Sub-Figures A-E), and the general anesthesia phase (Sub-Figures F-J). Interactions that increased are represented as continuous black arrows; interactions that decreased are represented as dashed red lines. Networks of the transition and general anesthesia phases were compared to those estimated during the resting state phase. Only the statistically significantly different connections compared to those respective to the resting state are represented. It is possible to notice that during the transition phase in the frequency bands Delta, Theta, Alpha, and Beta, the interactions from the occipital and temporal lobes toward the frontal and parietal lobe increased, which indicates a flow of Granger causality from inferior-posterior to anterior brain cortical areas (see Sub-Figures A, B, C, and D). During the transition phase, the statistical test did not confirm a decrease in functional connectivity among brain lobes. Under general anesthesia conditions, an increase in the interactions established among temporal and occipital lobes was verified in both directions (see Sub-Figures I-J). A decrease from the parietal to the occipital lobe was noticed in Delta Theta and Alpha frequency bands (see Sub-Figures F, G, and H).
}\label{fig:1}
\hypertarget{FIGURE4}{}
\end{figure}

\begin{multicols}{2}


\subsubsection{Functional Connectivity at Brain Lobes Level}

The functional connectivity at the brain lobe's level was evaluated to verify the alterations that happened in each brain lobe's functional connectivity due to propofol's administration. This measure considered all the connections in which nodes of the respective brain lobes participated and evaluated both intra-lobe connectivity (connections established among the nodes of the same lobe) and inter-lobe connectivity (connections established among the specific brain lobe to nodes of all other brain lobes).

\begin{itemize}

\item Transition Phase

In the Theta (4-8Hz), Alpha (8-12Hz), Beta (13-30Hz), and Gamma (25-100Hz) frequency bands, the Wilcoxon test confirmed that functional connectivity increased in all brain lobes analyzed (frontal, parietal, temporal, and occipital). In Delta (0-4Hz), the Wilcoxon test confirmed an increase in the frontal, temporal, and occipital lobes, while at the parietal lobe, the test was inconclusive.
These results reveal that during the transition phase, all brain lobes presented an increase in functional connectivity; there was no specific brain lobe in which functional connectivity decreased. 

The transition phase was essentially characterized by increased functional connectivity at the brain lobe's level.

\item General Anesthesia Phase

In the Alpha (8-12Hz) frequency band, functional connectivity decreased at the parietal lobe.
In the Beta (13-30Hz) frequency band, functional connectivity decreased at the frontal and temporal lobes while it increased at the occipital lobe.
In the Gamma (25-100Hz) frequency band, functional connectivity increased in the temporal and occipital lobes. 

Regarding functional connectivity at the brain lobes level, we have verified that the general anesthesia state was not solely characterized by a decrease in functional connectivity at all brain lobes; some brain lobes presented an increase in functional connectivity, for example, the occipital lobe in Beta (13-30Hz) and the occipital and temporal lobes in Gamma (25-100Hz).

\end{itemize}


\end{multicols}

\begin{figure}[!h]
\centering
  \begin{subfigure}[b]{.32\linewidth}
    \centering
    \includegraphics[width=.99\textwidth]{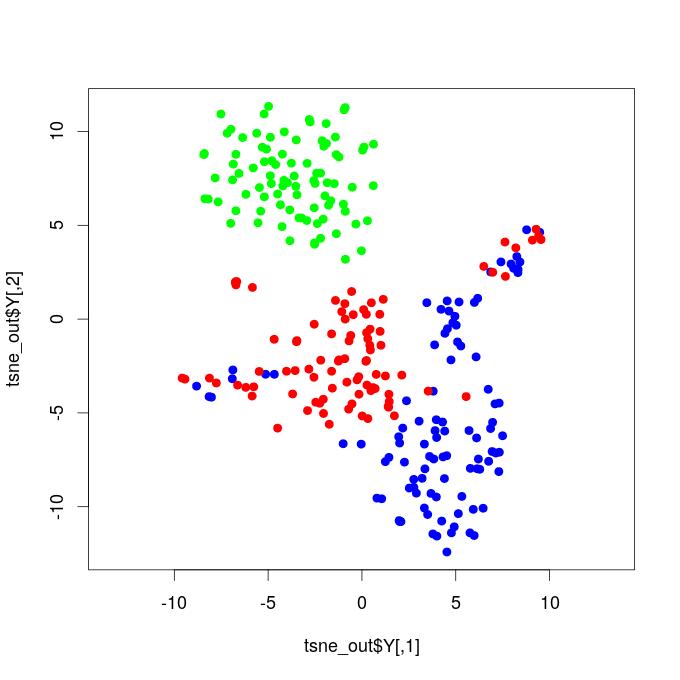}
    \caption{Delta (0-4Hz)}\label{fig:1a}
  \end{subfigure}%
  \begin{subfigure}[b]{.32\linewidth}
    \centering
    \includegraphics[width=.99\textwidth]{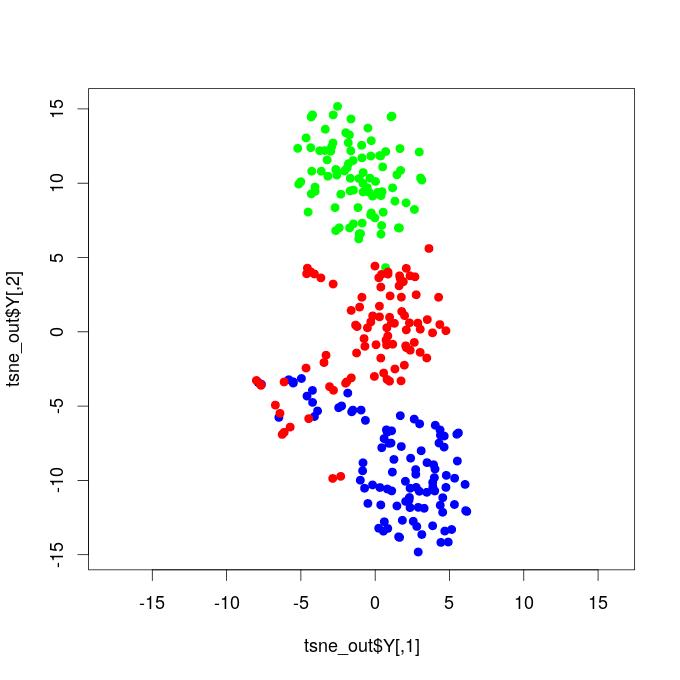}
    \caption{Theta (4-8Hz)}\label{fig:1b}
  \end{subfigure}%
  \begin{subfigure}[b]{.32\linewidth}
    \centering
    \includegraphics[width=.99\textwidth]{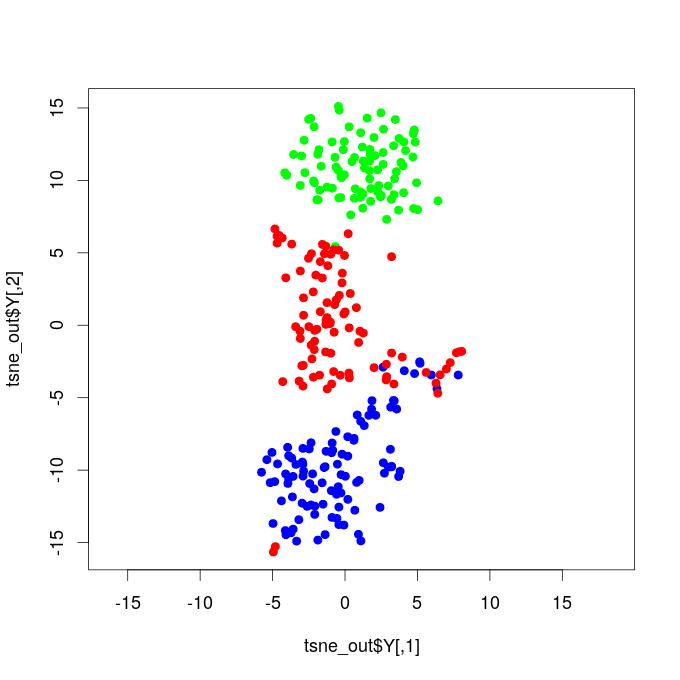}
    \caption{Alpha (8-12Hz)}\label{fig:1c}
  \end{subfigure}\\%
  \begin{subfigure}[b]{.32\linewidth}
    \centering
    \includegraphics[width=.99\textwidth]{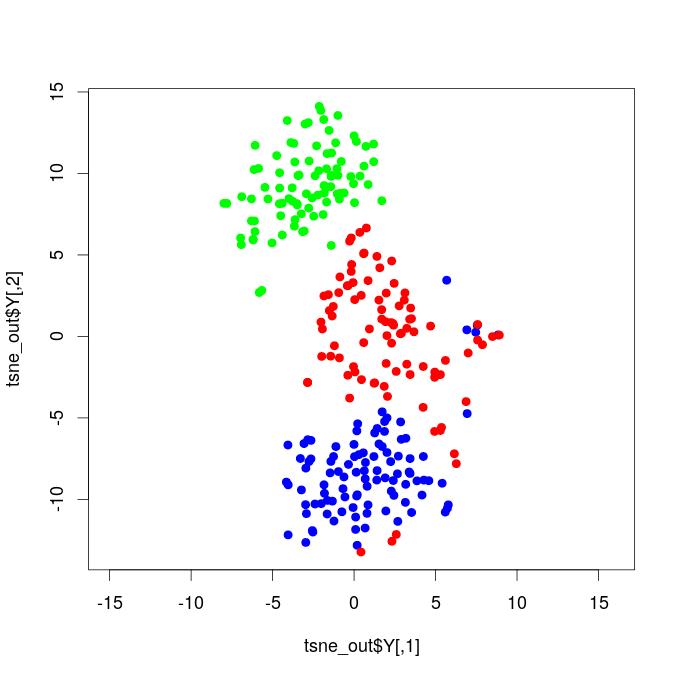}
    \caption{Beta (13-30Hz)}\label{fig:1d}
  \end{subfigure}%
  \begin{subfigure}[b]{.32\linewidth}
    \centering
    \includegraphics[width=.99\textwidth]{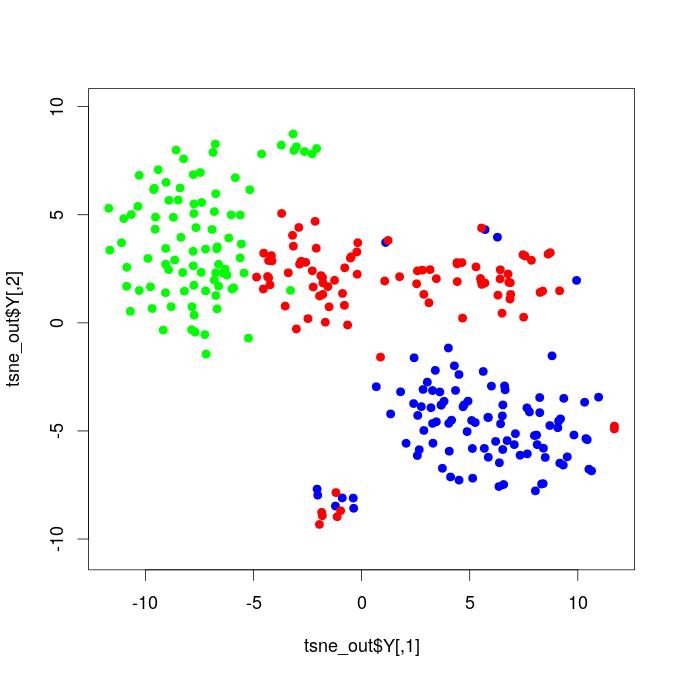}
    \caption{Gamma (25-100Hz)}\label{fig:1e}
  \end{subfigure}%
   \begin{subfigure}[b]{.32\linewidth}
    \centering
    \includegraphics[width=.40\textwidth]{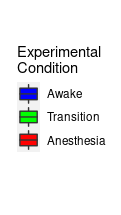}
      \end{subfigure}%
  \caption{\textbf{Projections of the time-resolved functional networks into a bi-dimensional space using a non-linear distance preserving t-SNE algorithm.} Functional cortical networks respective to the resting-state conditions are represented in blue, networks of the transition phase are represented in green, and networks corresponding to the general anesthesia conditions are indicated in red (see legend). It is possible to verify that networks corresponding to distinct stages of the experiment occupied particular regions of the bi-dimensional space. This result demonstrates that each state's functional cortical networks were distinct and specific. It can also be noticed that networks from the general anesthesia phase were closer to the networks of the resting state than those respective to the transition phase. From this observation, it can be concluded that the networks of general anesthesia are much more similar to those found during the awake-resting state conditions than those respective to the transition phase.
}\label{fig:1}
\end{figure}
\hypertarget{FIGURE5}{}
\begin{multicols}{2}


\vspace*{5mm}

\section{Discussion}

Studies demonstrated that specific mutations on the GABA-A receptors are able to eliminate or at least significantly weaken propofol anesthetic sensitivity \citep{belelli1997interaction,mihic1997sites} and concluded that GABA-A receptors are the main molecular target of propofol on the central nervous system, and the drug's pharmacological effects occur mainly due to the modulation of GABA-A receptors. The molecular mechanism of action has also been described. 
When propofol binds post-synaptically to GABA-A receptors, it potentiates inward chloride currents, which results in the hyperpolarization of the post-synaptic neurons and thus promotes inhibition \citep{bai1999general,hemmings2005emerging}. By enhancing gabaergic inhibition at the thalamic reticular nucleus, propofol decreases excitatory inputs from the thalamus to the cortex. However, these inhibitory effects of propofol do not lead to a complete brain inactivation but promote large numbers of brain circuits to act in a coordinated manner and lead to a state characterized by highly structured electrophysiological oscillatory patterns \citep{ching2014modeling}, which constitute a distinctive signature of propofol on the EEG.

By recording electrophysiological activity from both the thalamus (different higher-order thalamic nuclei) and the cortex (different layers of the medial prefrontal cortex) simultaneously in propofol anesthetized rats, \citep{flores2017thalamocortical} verified that the alpha oscillations are synchronized between the thalamus and the cortex, therefore evidenced that the thalamus and the cortex enter into an oscillatory coherent state \citep{flores2017thalamocortical}. This coordinated cortico-thalamic oscillatory state is often regarded as being correlated with anesthetic-induced unconsciousness conditions. However, it still needs to be determined how these consistently observed patterns are precisely related to mechanisms that promote unconsciousness under propofol-induced general anesthesia \citep{lewis2012rapid}.

 As long as the cortex and functional cortical networks are largely regarded as supporting neural processes that account for conscious experiences and cognitive capacity \citep{mcintosh2000towards,stam2007graph,tononi1998consciousness}, a better understanding of how the spatial and temporal organization of cortical dynamics evolves during the anesthetic induction is required in order to identify possible mechanisms by which propofol promotes unconsciousness.
 
By analyzing simultaneous local field potentials (LFPs) and ECoG records from temporal lobe cortical areas in human patients along with propofol anesthetic induction, \citep{lewis2012rapid} verified that significant structured correlation was maintained among nearby neurons, and spikes typically fluctuated over time. These results, therefore, revealed that, after LOC is established, tiny regions of the cortex have dynamic and coherent activity. 
By using a phase-locking factor (PLF), \citep{lewis2012rapid} quantified, on slow oscillations, the phase relationships among different cortical regions. It was found that distinct regions were often at different phases, and these phase differences were also not stable over time. The combination of strong local coupling of spikes and phase offsets among distant regions demonstrates that propofol was able to disrupt communication between distant cortical areas and thus affect functional connectivity \citep{lewis2012rapid}. Therefore, general anesthesia-induced unconsciousness supposedly arises as long as propofol disrupts cortical integration.

Some results and phenomena observed in our research with \textit{Macaca fuscata} are in accordance with the results presented in \citep{lewis2012rapid}. We observed that during general anesthesia, brain activity did not reduce to vanish completely. During general anesthesia, we contemplated that the brain still remained dynamic and active. We also verified that the administration of propofol in the macaque animal models was able to affect cortical functional connectivity and brain networks. 

In our research, the MDR-ECoG recording technique provided extensive coverage of cortical brain areas, and we could contemplate the effects of propofol almost all over the cortical surface. We have found that the state of general anesthesia was not solely given due to an overall decrease in functional connectivity throughout the entire cortex. While functional interactions established among some cortical regions were affected, there were still some specific areas in which functional connectivity, in fact, increased; these results, therefore, reveal that the effects of propofol over the cortex are diverse along with specialized cortical areas.\\

Another study evaluated directed functional connectivity among large areas of the cortex. By analyzing multichannel electroencephalography records from frontal and parietal scalp areas in human patients, \citep{lee2009directionality} applied a \textit{cross-dependence analysis} \citep{rosenblum2001detecting} among the EEG signals during awake and propofol-induced anesthesia conditions. By doing so, \citep{lee2009directionality} evaluated feedforward \citep{lee2009directionality} [parietal $ \rightarrow $ frontal] and feedback \citep{lee2009directionality} (frontal $ \rightarrow $ parietal) functional connectivity interactions. The researchers \citep{lee2009directionality} verified that during propofol general anesthesia, feedforward [parietal$ \rightarrow $ frontal] communication was essentially maintained. In contrast, feedforward [frontal $ \rightarrow $ parietal] communication was compromised, existing an asymmetry in which the functional connectivity was predominantly affected in just one direction.

In our study, we also observed phenomena with similar characteristics as reported by \citep{lee2009directionality} in human patients. We also have found asymmetry in the functional connectivity directions established among brain lobes. The research conducted by \citep{lee2009directionality} was restricted to neural records from frontal and parietal areas. Our study contemplated a much more extensive coverage of the macaque's animal model cortical surface, as electrophysiological activity was recorded from the frontal, parietal, temporal, and occipital lobes lateral surfaces and also in the frontal, parietal, and medial occipital walls.\\

 We have observed that feedforward functional connectivity (from the parietal toward the occipital lobe) was significantly higher during general anesthesia than during resting state, on frequency bands Theta and Alpha. Although we did not find statistically significant alterations in the feedback (from frontal toward parietal lobe) functional connectivity precisely as verified by \citep{lee2009directionality}, we have found that on the Delta and Alpha frequency bands, functional connectivity from parietal to occipital lobe has been compromised, and verified that functional connectivity in the backward direction (from frontal and parietal lobes toward occipital and temporal lobes) were most likely to be compromised during general anesthesia.

We have used a Granger causality methodology to analyze a neural records database from \textit{Macaca fusctata} animal models, performed with the MDR-ECoG technique, which offered simultaneously high spatial coverage along with high temporal and spatial resolution. Our research complements the actual scenery and brings novel knowledge toward the characterization of the propofol effects on cortical functional connectivity along the anesthetic induction process.
 The present research is one of the first studies to estimate time-resolved functional networks serially over time to follow the unfolding of propofol anesthetic induction.

We verified that after the drug administration and before LOC was achieved, a transition state demonstrated to be quite particular from the general anesthesia state. This transition state was characterized by increased overall cortical functional connectivity in conjunction with a specific Granger causality flow profile among the cortical lobes. During the transition phase, it was mainly observed that a Granger causality flow from temporal and occipital areas, which assumed the character of a source of Granger causality, towards frontal and parietal regions, which figured the character of a sink of Granger causality.

In addition, our research also brings new knowledge regarding the general anesthesia state. It was verified that an increase in the functional connectivity among the electrodes located over the occipital lobe on the five frequency bands analyzed presented an increase in the connectivity at the local level in the occipital lobe. 
During general anesthesia, it was also verified that an increase in functional connectivity was established among the occipital and temporal lobes in both directions at higher frequency bands (both Beta and Gamma). Furthermore, an augment in the interactions from the parietal to the frontal lobe occurred on the frequency bands Theta and Alpha. Conversely, the connectivity from the parietal toward the occipital lobe decreased in the Delta, Theta, and Alpha frequency bands, showing that functional connectivity from frontal and parietal areas to occipital and temporal cortical areas had been compromised.
These results reveal that the effects of propofol along the cortex are diverse, and the general anesthesia state does not strictly follow an overall decrease in functional connectivity along the entire cortex. While propofol, in fact, can compromise communication in some neuronal pathways, some circuits might also have been activated. These results reinforce statements that claim propofol-induced general anesthesia as a particular and specific state.

\subsection{T-SNE Projections - Functional Brain Networks Three Distinct States}

\enlargethispage{-1.5\baselineskip}

Our results demonstrate that propofol-induced anesthesia is not given through a complete cortical functional network failure. It was verified that during general anesthesia, the brain keeps active and cortical areas still functionally interact. Compared to the resting state networks, the statistical test did not confirm that the mean functional connectivity of general anesthesia networks increased or decreased at global levels. This result reveals that propofol-induced unconsciousness is not necessarily given by means of a significant reduction in global functional connectivity. We projected the time-resolved functional networks into a bi-dimensional space using a non-linear distance preserving t-SNE algorithm \citep{van2008visualizing}. We then verified that populations of networks respective to the resting, transition, and general anesthesia states were located in particular regions of the bi-dimensional space, being reasonably separated, thus revealing that brain activity is distinct and specific during each state. This result does not go in accordance with hypotheses that, during general anesthesia, brain functional networks should be similar to networks respective to the resting state, typically resembling the structural brain networks that underlie functional connectivity. Instead, our results befit that functional networks under general anesthesia are characteristic and specific; certain connections were affected, while others might have been favored and configured in a reasonably distinct state from the brain's natural resting conditions. By analyzing the t-SNE projections, it was also noticed that general anesthesia functional networks were displayed closer to the resting state networks than to those of the transition state; this outcome might be due to the mean functional connectivity, which did not vary too much among these two states, while it remarkably increased during the transition.

\subsubsection{Propofol-Induced General Anesthesia  Vs Ketamine-Medetomidine Induced General Anesthesia}

General anesthesia is a drug-induced state that compromises unconsciousness, amnesia, analgesia, and immobility along with the maintenance of physiological stability \citep{brown2010general,brown2011general}. A series of drugs can induce physiological states that involve unconsciousness \citep{franks2008general}. The hypothesis of whether different anesthetic agents shared one common mechanism and the same mode of action in producing unconsciousness once has been the subject of much debate \citep{angel1993central}. Today, it is appreciated that distinct anesthetic drugs produce anesthetic states that might differ appreciably from one another. As Ketamine and Medetomidine, which act in the same brain circuits as propofol \citep{ching2014modeling}, however, modulate in particular NMDA and alpha-2-adrenergic receptors, respectively, and can induce quite distinctive brain dynamics compared to propofol \citep{brown2010general,brown2011general}. Other studies performed with the same macaque animal models, MDR-ECoG recording technique, and Granger causality in the frequency domain methodology \citep{padovani2016characterization,padovani2016structure}, reported effects on brain dynamics that occurred along with the administration of a Ketamine-Medetomidine cocktail. The effects were substantially different from those observed in the present research induced by propofol.

 The transition of Ketamine-Medetomidine anesthetic induction started within about 30 seconds of the injection of the anesthetic agents. It was characterized by a remarkable decrease in functional connectivity, which abruptly occurred in a similar single-step process. On the other hand, the propofol anesthetic induction transition was essentially patterned by an increase in functional connectivity and followed a more complex unfolding, in which functional connectivity gradually increased until a maximum was achieved and then decreased progressively until the mean LOC functional connectivity was reached. 


The Ketamine-Medetomidine general anesthesia-induced state was characterized by a significant reduction in functional connectivity that occurred mainly over areas of the associational cortex. While in the propofol general anesthesia-induced state, the mean brain functional connectivity has not been drastically reduced, despite communication established among certain cortical lobes being compromised, configuring a fairly distinct state from the state produced by Ketamine-Medetomidine. One common feature shared by both propofol and Ketamine-Medetomidine general anesthesia states was increased functional connectivity at a local level in the occipital lobe.

\subsubsection{Similarity and Dissimilarity Among the Experiments and Subjects}

In general, administering propofol resulted in a consistent dynamic response behavior, which was reproducible in the distinct experiments and over both animal models; however, some differences were also observed. 

Experiments involving the same animal model were quite similar to each other; the t-SNE projections of the connectivity matrices displayed similar patterns for the same individuals, as did the functional connectivity patterns (see complementary material). However, some distinctions among the experiments also occurred (see complementary material) and are supposedly due to the intrinsic variability of functional connectivity.

More significant variability was observed across individuals, the macaque A displayed a more extended transition period than the macaque B. Differences in the mean connectivity, mainly at the brain lobe's level, and in the Granger causality flow among brain lobes were also verified among subjects (see complementary material). These dissimilarities might have come from distinctions of individual characteristics such as size, weight, and age, as well as differences in the experimental setup; as reported by \citep{wang2012memory} propofol anesthetic induction might present changes according to the subject's individual characteristics. The position of the ECoG electrode matrix over the left brain cortical surface was just slightly different. However, electrodes were positioned in a relatively distinct manner over the medial cortical walls; macaque A had eight electrodes placed at the medial occipital walls and no electrodes placed at the medial parietal walls, and macaque B had eight electrodes positioned at medial parietal walls and no electrodes placed at the medial occipital walls. As the positions of the electrodes were not the same, dissimilarities among the individuals were expected by default.

\enlargethispage{\baselineskip}

\subsection{Conclusions}

With the MDR-ECoG recording technique and Granger causality in the frequency domain methodology, we could contemplate the unfolding of the propofol anesthetic induction in the macaque's functional brain networks. The most remarkable phenomenon observed in the research was an expressive increase in functional connectivity, which occurred after the anesthetic drug administration during the transition phase. This state was also characterized by a specific Granger causality flow profile, from the occipital and temporal to frontal-parietal cortical lobes. It was also verified that during general anesthesia, the magnitude of the mean functional connectivity was about the same as during the resting state. However, the functional connections established among certain brain regions had been compromised while others had been favored, verifying alterations in the five frequency bands analyzed. The present research complements the actual scenery and brings novel knowledge toward the characterization of the functional connectivity dynamics throughout propofol anesthetic induction in non-human primates.

The exact mechanisms of general anesthesia are still incompletely known, and the characterization of functional connectivity in the anesthetized states induced by different drugs is, for sure, one essential step for the comprehension of neural correlates of general anesthesia and the elucidation of neuronal pathways and circuits that are involved with unconsciousness, amnesia, and immobility. This elucidation could help in the design of novel drugs or a combination of drugs, that could promote a better and safer anesthetic induction that would best fit the needs of clinical applications. It could also offer new knowledge about mechanisms that involve the depression of the central nervous system, such as sleep. The characterization of functional connectivity during general anesthesia induced by distinct anesthetic agents can also constitute novel relevant knowledge for neuroscience, as it may help the comprehension of neural correlates of consciousness and offer more insights into essential neural processes that account for the brain to perform its activities.

\enlargethispage{5\baselineskip}

\bibliography{bibfile2}

\end{multicols}

\end{document}